\documentclass[fleqn,usenatbib,useAMS]{mnras}

\usepackage{newtxtext,newtxmath}
\usepackage[T1]{fontenc}
\DeclareRobustCommand{\VAN}[3]{#2}
\let\VANthebibliography\thebibliography
\def\thebibliography{\DeclareRobustCommand{\VAN}[3]{##3}\VANthebibliography}
\usepackage{pdflscape}	
\usepackage{ae,aecompl}

\usepackage{graphicx} 

\newcommand{\clfd}{\mbox{\textsc{clfd}}}

\newcommand{\presto}{\mbox{\textsc{presto}}}

\newcommand{\psrchive}{\mbox{\textsc{psrchive}}}
\newcommand{\pulsarX}{\mbox{\textsc{pulsarX}}}
\newcommand{\psrfoldfil}{\mbox{\texttt{psrfold\_fil}}}
\newcommand{\filtool}{\mbox{\texttt{filtool}}}
\newcommand{\candyjar}{\mbox{\textsc{CandyJar}}}
\newcommand{\peasoup}{\mbox{\textsc{peasoup}}}
\newcommand{\pics}{\mbox{\textsc{pics}}}
\newcommand{\iqrm}{\mbox{\textsc{iqrm}}}
\newcommand{\dedisp}{\mbox{\textsc{dedisp}}}
\newcommand{\mosaic}{\mbox{\textsc{mosaic}}}
\newcommand{\transientX}{\mbox{\textsc{transientX}}}

\newcommand{\multiTRAPUM}{\mbox{\textsc{multiTRAPUM}}}
\newcommand{\aplpy}{\mbox{\textsc{aplpy}}}
\newcommand{\pygsm}{\mbox{\textsc{pyGSM}}}

\newcommand{\dmunits}{\,pc\,cm$^{-3}$}

\newcommand{\h}{$^{\rm h}$} 
\newcommand{\m}{$^{\rm m}$}

\newcommand{\SextansAposition}{\mbox{RA(J2000) $=$ 10\h11\m00\fs5} \mbox{Dec(J2000) $=$ $-$04\textdegree{}41\arcmin30\farcs0}}

\newcommand{\SextansBposition}{\mbox{RA(J2000) $=$ 10\h00\m00\fs82} \mbox{Dec(J2000) $=$ 05\textdegree{}20\arcmin09\farcs57}}

\title[TRAPUM pulsar search of Sextans A and B]{TRAPUM pulsar and transient search in the Sextans A and B galaxies and discovery of background FRB\,20210924D}

\author[E. Carli et al.]{\parbox{\textwidth}{
E. Carli,$^{1}$\thanks{E-mail: \href{mailto:emma.carli@outlook.com}{emma.carli@outlook.com}}
L. Levin,$^{1}$
B. W. Stappers,$^{1}$
E. D. Barr,$^{2}$
R. P. Breton,$^{1}$
S. Buchner,$^{3}$
M. Burgay,$^{4}$
M. Kramer,$^{2}$
P. V. Padmanabh,$^{5,6,2}$
A. Possenti,$^{4}$
V. Venkatraman Krishnan,$^{2}$
S. S. Sridhar,$^{7,2}$
J. D. Turner$^{1}$
}
\\ \\ \\
$^{1}$Jodrell Bank Centre for Astrophysics, Department of Physics and Astronomy, The University of Manchester, Manchester M13 9PL, UK \\
$^{2}$Max-Planck-Institut f\"{u}r Radioastronomie, Auf dem H\"{u}gel 69, D-53121 Bonn, Germany \\
$^{3}$South African Radio Astronomy Observatory (SARAO), 2 Fir Street, Black River Park, Observatory, Cape Town, 7925 \\
$^{4}$INAF-Osservatorio Astronomico di Cagliari, via della Scienza 5, 09047, Selargius, Italy \\
$^{5}$Max-Planck-Institut f\"{u}r Gravitationsphysik (Albert-Einstein-Institut), D-30167 Hannover, Germany\\
$^{6}$Leibniz Universit\"{a}t Hannover, D-30167 Hannover, Germany\\
$^{7}$SKA Observatory, Jodrell Bank, Lower Withington, Macclesfield SK11 9FT, UK\\
}

\date{Accepted XXX. Received YYY; in original form ZZZ}
\pubyear{2024}

\begin{document}
\label{firstpage}
\pagerange{\pageref{firstpage}--\pageref{lastpage}}
\maketitle

\begin{abstract}
The Small and Large Magellanic Clouds are the only galaxies outside our own in which radio pulsars have been discovered to date. The sensitivity of the MeerKAT radio interferometer offers an opportunity to search for a population of more distant extragalactic pulsars.
The TRAPUM (TRansients And PUlsars with MeerKAT) collaboration has performed a radio-domain search for pulsars and transients in the dwarf star-forming galaxies Sextans A and B, situated at the edge of the local group 1.4\,Mpc away. 
We conducted three 2-hour multi-beam observations at L-band (856--1712\,MHz) with the full array of  MeerKAT.
No pulsars were found down to a radio pseudo-luminosity upper limit of 7.9$\pm$0.4\,Jy\,kpc$^{2}$ at 1400\,MHz, which is 28 times more sensitive than the previous limit from the Murriyang telescope. This luminosity is 30\,per cent greater than that of the brightest known radio pulsar and sets a cut-off on the luminosity distributions of the entire Sextans A and B galaxies for unobscured radio pulsars beamed in our direction.
A Fast Radio Burst was detected in one of the Sextans A observations at a Dispersion Measure (DM) of 737\dmunits{}.  We believe this is a background event  not associated with the dwarf galaxy due to its large DM and its S/N being strongest in the wide-field incoherent beam of MeerKAT.

\end{abstract}

\begin{keywords}
stars: neutron -- pulsars: general -- galaxies: individual: Sextans A, Sextans B --  transients: fast radio bursts  -- radio continuum: transients -- galaxies: Local Group
\end{keywords}
\section{Introduction}
\label{introduction}
The Small and Large Magellanic Clouds (SMC and LMC) are the only galaxies outside the Milky Way in which radio pulsars have been discovered to date  (\citealt{Hisano2022}, \citealt{Titus2019}, and references therein) due to their proximity and position away from the Galactic plane.  Of the nearly 3400 radio pulsars that have been discovered, only 31 are extragalactic (see the \href{https://www.atnf.csiro.au/research/pulsar/psrcat/proc_form.php?version=2.1.1&Name=Name&JName=JName&RaJ=RaJ&DecJ=DecJ&P0=P0&P1=P1&F2=F2&DM=DM&RM=RM&W50=W50&Tau_sc=Tau_sc&S1400=S1400&Dist=Dist&Assoc=Assoc&Date=Date&Type=Type&R_lum14=R_lum14&Age=Age&Bsurf=Bsurf&Edot=Edot&startUserDefined=true&c1_val=&c2_val=&c3_val=&c4_val=&sort_attr=jname&sort_order=asc&condition=Dist+%3E+40&pulsar_names=&ephemeris=short&coords_unit=raj%2Fdecj&radius=&coords_1=&coords_2=&style=Long+with+last+digit+error&no_value=*&fsize=3&x_axis=&x_scale=linear&y_axis=&y_scale=linear&state=query&table_bottom.x=74&table_bottom.y=16}{Australia Telescope National Facility (ATNF) pulsar catalogue} for a list, \citealt{ATNF}).  So far, no rotation-powered pulsars have been found at distances larger than that of the SMC \citep[60\,kpc,][]{Karachentsev2004}.
Extragalactic pulsars are sought after for insights on the influence of different galactic properties on neutron star (NS) formation. For example, the impact of galactic metallicity and star formation history on stellar mass and supernova physics can be investigated \citep{Heger2003,Titus2020}. A rotation-powered pulsar outside the Magellanic Clouds would extend these studies to a new galaxy.

Sextans A and B are dwarf irregular faint galaxies at the edge of the local group, 1.4\,Mpc away \citep{Dalcanton2009}, and no pulsars are known within three degrees of their location \citep{ATNF}.  They are 260\,kpc away from each other \citep{Dohm-Palmer1997},  with sizes of about 4\,kpc \citep{Bellazzini2014}.
The expected Milky Way Dispersion Measure (DM) contribution in their direction is low: about $32$ or $45$\dmunits{}  according to the YMW2016 or NE2001 electron density models respectively \citep{YMW2016,NE2001}. This is of the same order of magnitude as the DM contributions in the direction of the Magellanic Clouds and is favourable to radio pulsar searches.
    
The galaxies are undergoing episodes of more intense star formation, which started 50\,Myr ago in Sextans A \citep{Dohm-Palmer1997} and 1--2\,Gyr ago in Sextans B \citep{sakai1997}.
Both are still star forming at the rate of 2--8 $\times 10^{-3} \text{M}_{\sun}$\,yr$^{-1}$ \citep{Weisz2011}.
This can lead to the presence of young pulsars like it has in the Magellanic Clouds\footnote{There was a recent episode of stellar formation in the SMC about 40\,Myr ago \citep{Harris2004}, and in the LMC 12\,Myr ago \citep{Harris2009}.} \citep[e.g.][]{Carli2024,Maitra2021,Lamb2002,Seward1984}, and an abundance of young systems per unit mass compared to the Milky Way \citep[e.g][]{Maggi2019}.
The discovery of young pulsars is scientifically interesting. Among them, magnetars (which have the highest known stellar magnetic fields) are important for a variety of fundamental physics and astrophysics investigations \citep{Esposito2021}, including as one of the progenitors of Fast Radio Bursts  \citep[FRBs,][]{Bochenek2020b, Platts2019}. One extragalactic magnetar has been discovered in each of the Magellanic Clouds \citep{Lamb2002,Helfand1979}; giant magnetar flares have been observed in nearby galaxies \citep{Trigg2023,Mereghetti2024}; and an FRB repeater has been found in a dwarf galaxy \citep{Tendulkar2017}: the powerful emission from these exotic systems can be detected at extragalactic distances. Additionally, young pulsars can emit giant narrow pulses \citep[e.g. the extragalactic Crab pulsar twin in the LMC,][]{Johnston2003}. If a pulsar is too distant and thus too faint to be detected even in a long integration, single pulses that are orders of magnitude stronger than average may still be detected individually, although  distance is still a challenge (see \autoref{discussion}). Sextans A and B host multiple shell-like structures formed as the result of photoionisation from massive stars, stellar winds, and supernovae. Each galaxy contains three candidate supernova remnants \citep{Gerasimov2022,Gerasimov2024}. There is also one supernova remnant identified in X-rays in Sextans A \citep{Kappes2005}. These can potentially host  young pulsars.

Low metallicity can strongly decrease stellar mass-loss rates through line-driven winds in progenitor stars and thus increase the mass of the final remnant. This can lead to an increased neutron star birth rate, and, possibly, increased NS masses \citep{Heger2003}. Sextans A and B have a lower metallicity\footnote{The mean [Fe/H] values are, from lowest  to highest metallicity: $-1.6$ for Sextans B, $-1.5$ for Sextans A \citep{Bellazzini2014},  $-0.7$ for the SMC, $-0.3$ for the LMC \citep{Luck1998}, and about 0 for the Milky Way \citep[e.g.][]{Hayden2014}.} than the Magellanic Clouds \citep{Garcia2019} and the Milky Way, therefore many scientifically valuable compact objects might be formed \citep[e.g.][]{Titus2020}. 
For example, the low-metallicity SMC hosts a 50 times higher density of High-Mass X-ray binary systems (a high-mass star in a binary with a NS or black hole) than in the Milky Way \citep{Haberl2016,Yang2017}.
High-mass systems that have not yet reached their final evolutionary stages are known in Sextans A under the form of a large population of early O-type stars \citep{Garcia2019,lorenzo2022} and accreting X-ray binaries \citep{Kappes2005}. This is indicative of systems that will form or contain neutron stars and recycled pulsars.
    
Finally, old star clusters are a prime target to search for millisecond (MSPs) and binary pulsars \cite[see][]{Lorimer2008}, as well as transients: an FRB was pinpointed to a globular cluster 3.7\,Mpc away in the galaxy M81 (twice the distance to the Sextans galaxies), which was until then an ill-favoured progenitor environment \citep{Bhardwaj2021,kirsten2022, kremer2023}. A low-metallicity, 9\,Gyr Globular Cluster (GC) is already known in Sextans A \citep{Beasley2019,Gvozdenko2024}. Besides this, Sextans B hosts a 1-3 Gyr star cluster \citep{Sharina2007}. Both these clusters could potentially harbour multiple millisecond pulsars. 

TRAPUM (TRAnsients and PUlsars with MeerKAT) is a Large Survey Project of the MeerKAT telescope (\href{http://trapum.org/}{trapum.org}, \citealt{Stappers2016}).  The collaboration has already discovered over 200 pulsars\footnote{\href{http://trapum.org/discoveries/}{http://trapum.org/discoveries/}}.  One of TRAPUM's science goals is to find new extragalactic pulsars. Motivated by the sensitivity of the MeerKAT telescope to faint, distant pulsars (as evidenced by recent discoveries in the SMC, see \citealt{Carli2024}) and new transients, we have conducted observations of the Sextans A and B galaxies. The observations, described in \autoref{observations}, encompass the whole galaxies and thus their entire pulsar population with favourable viewing geometries, which as detailed above could be rich in interesting systems. We detail our search analysis in \autoref{search}, and report the detection of a  Fast Radio Burst in \autoref{FRB}, likely originating from outside the targeted dwarf galaxy. We state our upper limits on radio pulsations from these galaxies in \autoref{upperlimits}. Considering the distance to these galaxies being 23 times larger than the distance to the Magellanic Clouds, we discuss what systems could be detected in \autoref{discussion}.

\section{Observations}
\label{observations}

\begin{table*}

\centering
	\caption{The Sextans A observation parameters. For each observation, a tiling of coherent beams is placed at the centre of the pointing, within the MeerKAT primary incoherent beam. The pointing is aimed at the centre of the Sextans A galaxy: \SextansAposition{}  \protect\citep{Paturel2003}. The coherent beam semi axes, fitted by \mosaic{}, are given in arcseconds in the Right Ascension and Declination directions, with the ellipse position angle (taken clockwise from North). }
	\label{tab:SextansA_observations}

\resizebox{\textwidth}{!}{

\begin{tabular}{ccccccc}
\textbf{Date}           & \textbf{Observation length} & \textbf{\begin{tabular}[c]{@{}c@{}}Number of \\ incoherent antennas\end{tabular}} & \textbf{\begin{tabular}[c]{@{}c@{}}Number of \\ coherent antennas\end{tabular}} & \textbf{\begin{tabular}[c]{@{}c@{}}Number of \\ coherent beams\end{tabular}} & \textbf{\begin{tabular}[c]{@{}c@{}}Coherent beam tiling\\ sensitivity overlap\end{tabular}} & \textbf{\begin{tabular}[c]{@{}c@{}}Coherent beam size \\ (50\% sensitivity)\end{tabular}} \\ \hline
\textbf{June 2021}      & 6798\,s                     & 57                                                                                & 56                                                                              & 246                                                                          & 47\%                                                                                        & \begin{tabular}[c]{@{}c@{}}26\farcs2 9\farcs9\\ -34\fdg1\end{tabular}                       \\ \hline
\textbf{July 2021}      & 7167\,s                     & 62                                                                                & 60                                                                              & 260                                                                          & 50\%                                                                                        & \begin{tabular}[c]{@{}c@{}}15\farcs0 10\farcs1\\  55\fdg6\end{tabular}                       \\ \hline
\textbf{September 2021} & 7085\,s                     & 62                                                                                & 60                                                                              & 260                                                                          & 50\%                                                                                        & \begin{tabular}[c]{@{}c@{}}19\farcs6 12\farcs3\\ 62\fdg9\end{tabular}                       \\ \hline
\end{tabular}

}
\end{table*}

\begin{table*}

\centering
	\caption{The Sextans B observation parameters. For each observation, a tiling of coherent beams was placed at the centre of the pointing, within the MeerKAT primary incoherent beam. The pointing was aimed at the centre of the Sextans B galaxy: \SextansBposition. The coherent beam major axes, fitted by \mosaic{}, are given in arcseconds in the Right Ascension and Declination directions, with the ellipse position angle (taken clockwise from North). The overlaps obtained may be different than intended due to updates to FBFUSE and the \mosaic{} software, see \protect\cite{Carli2024} for more details. }
	\label{tab:SextansB_observations}

\resizebox{\textwidth}{!}{
\begin{tabular}{ccccccc}
\textbf{Date}           & \textbf{Observation length} & \textbf{\begin{tabular}[c]{@{}c@{}}Number of \\ incoherent antennas\end{tabular}} & \textbf{\begin{tabular}[c]{@{}c@{}}Number of \\ coherent antennas\end{tabular}} & \textbf{\begin{tabular}[c]{@{}c@{}}Number of \\ coherent beams\end{tabular}} & \textbf{\begin{tabular}[c]{@{}c@{}}Coherent beam tiling\\ sensitivity overlap\end{tabular}} & \textbf{\begin{tabular}[c]{@{}c@{}}Coherent beam size \\ (50\% sensitivity)\end{tabular}} \\ \hline
\textbf{June 2021}      & 7156\,s                     & 57                                                                                & 56                                                                              & 266                                                                          & 50\%                                                                                        & \begin{tabular}[c]{@{}c@{}}17\farcs9 10\farcs8\\ -39\fdg1\end{tabular}                       \\ \hline
\textbf{July 2021}      & 7152\,s                     & 62                                                                                & 60                                                                              & 252                                                                          & 50\%                                                                                        & \begin{tabular}[c]{@{}c@{}}14\farcs9 8\farcs2\\ 76\fdg1\end{tabular}                         \\ \hline
\textbf{September 2021} & 6540\,s                     & 62                                                                                & 60                                                                              & 252                                                                          & 50\%                                                                                        & \begin{tabular}[c]{@{}c@{}}19\farcs6 9\farcs9\\ 76\fdg2\end{tabular}                       \\ \hline
\end{tabular}

}
\end{table*}

\begin{figure*}
\centering
\includegraphics[width=\textwidth]{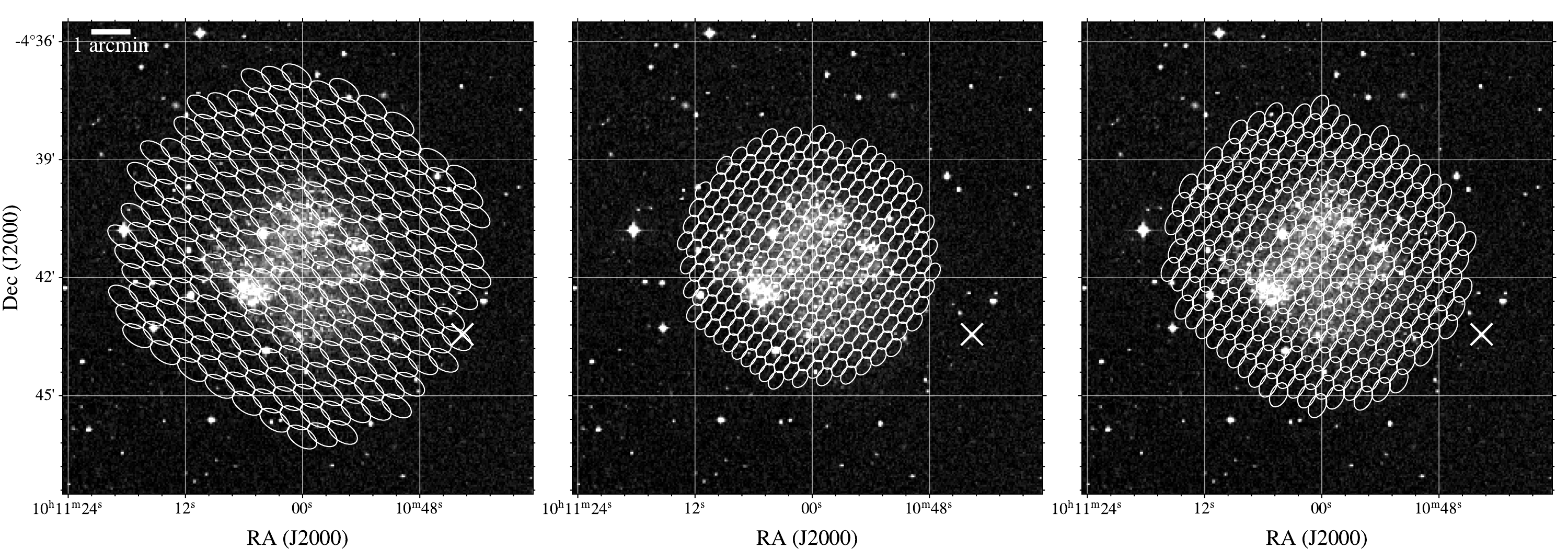}
\caption[caption:SextansA-tilings]{The three observations of Sextans A (first to third pass left to right) are displayed over a DSS image \citep{DSS}. The observed MeerKAT full array coherent beams are shown at 50 per cent sensitivity (47 per cent for the first pass) as white ellipses, simulated with \mosaic{}. The MeerKAT incoherent beam size is larger than the figure with a FWHM of about 1\,degree at L-band. This figure was generated with the \href{http://aplpy.github.io/}{\aplpy{}} Python package. The Globular Cluster \citep{Beasley2019} is shown with a cross.  The beam orientation changes are due to the source being observed when rising in the first pass and setting in the two other passes. The beam size depends on the Hour Angle of the source and the location of the antennas used.}
\label{fig:SextansA-tilings}
\end{figure*}

\begin{figure*}
\centering
\includegraphics[width=\textwidth]{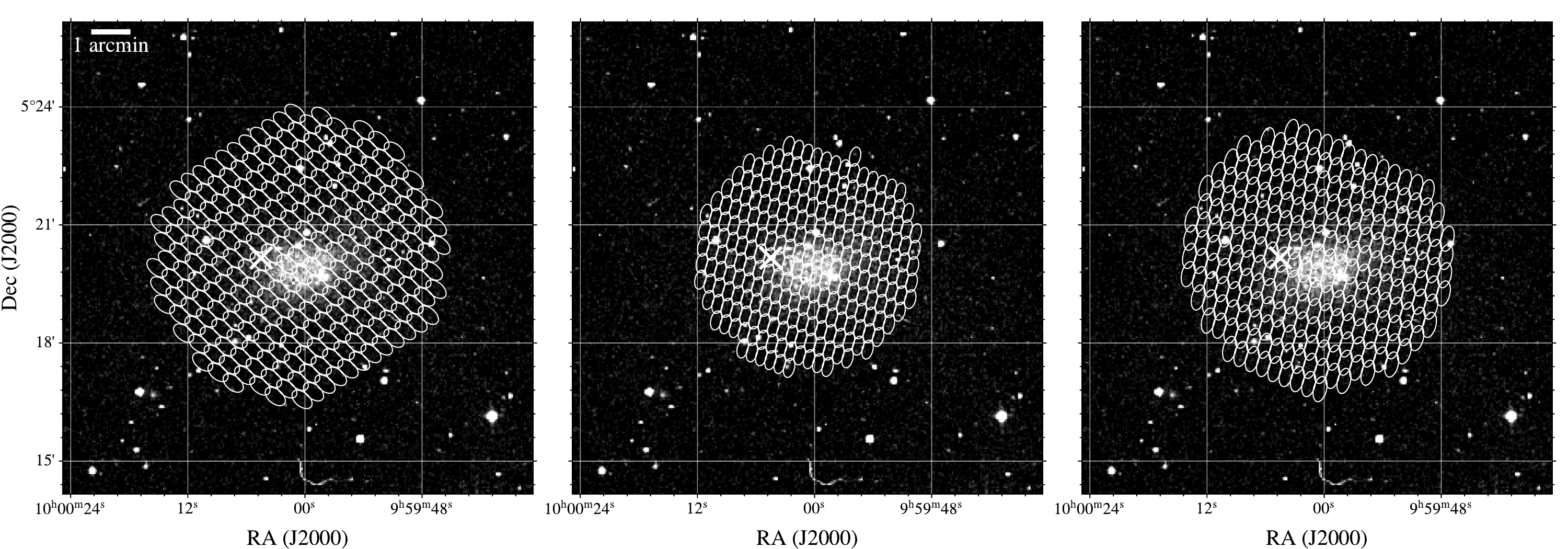}
\caption[caption:SextansB-tilings]{The three observations of Sextans B (first to third pass left to right) are displayed over a DSS image. The MeerKAT incoherent beam size is larger than the figure. The observed MeerKAT full array coherent beams are shown at 50 per cent sensitivity as white ellipses, simulated with \mosaic{}. This figure was generated with the \href{http://aplpy.github.io/}{\aplpy{}} Python package. The star cluster \citep{Sharina2007} is shown with a cross. The beam orientation changes are due to the source being observed when rising in the first pass and setting in the two other passes. The beam size depends on the Hour Angle of the source and the location of the antennas used.}
\label{fig:SextansB-tilings}
\end{figure*}

A detailed introduction to TRAPUM's observing strategy for nearby galaxies with MeerKAT is presented in \cite{Carli2024}. We conducted three 2-hour multi-beam observations of the dwarf galaxies Sextans A and B. For each observation, a tiling of coherent beams was placed at the centre of the pointing (i.e. at the centre of the MeerKAT primary incoherent beam that has a Full Width at Half-Maximum of about 1\,degree at L-band). The pointings were aimed at the centre of each  galaxy.  We used the TRAPUM backends Filterbanking Beamformer User Supplied Equipment (FBFUSE) and Accelerated Pulsar Search User Supplied Equipment (APSUSE) to synthesise and record about 260 beams as filterbank search-mode files \citep{Padmanabh2023,Chen2021}. We carried out the observations at L-band (856--1712\,MHz), with 2048 frequency channels and a sampling time of 153\,$\upmu$s. The beams were formed using the MeerKAT full array (56 to 60 antennas out of 64 due to availability) with a maximum baseline length of about 8\,km. 
Sextans A and B are small sources, with a diameter of about 5--6 arcminutes  \citep{Bellazzini2014}. Despite the much reduced beam size of the full array of MeerKAT compared to its core configuration, we were able to cover each entire galaxy with about 260 beams and comply with APSUSE data rate recording requirements. This allowed us to obtain the best sensitivity for our integration time. The characteristics of the individual observations are given in \autoref{tab:SextansA_observations}  and \autoref{tab:SextansB_observations}. In \autoref{fig:SextansA-tilings} and \autoref{fig:SextansB-tilings}, we show the layout of the MeerKAT coherent beams for each observation. The beam positions were retrieved from the FBFUSE record of the observation. A high resolution coherent beam Point Spread Function (PSF)  was simulated with \mosaic{} \citep{Chen2021} at the centre coordinates of the pointing, at the central time of the observation, with the antennas used during the observation and at the central frequency of L-band. The ellipse-shaped beam contours, fitted by \mosaic{} to the PSF, are shown at the sensitivity overlap of the tiling.  One 2-hour pass on both sources occupies around 50 terabytes of disk space on the APSUSE cluster.

Imaging data were recorded commensally for each observation as correlated visibilities with the full array. The visibility data were correlated in the `wideband coarse (4k)' mode, resulting in 4096 208.98 kHz-wide channels to cover the 856--1712 MHz frequency range. The time resolution of the visibility data is 8s.

\section{Data reduction}
\label{search}

The data were processed locally, on the APSUSE computing cluster via the standard TRAPUM search pipeline (see Section 6.1 in \citealt{Padmanabh2023}). We first processed the survey data with a periodicity search pipeline.
Radio Frequency Interference (RFI) cleaning was performed for the first two passes with a combination of \iqrm{} \citep{IQRM}, and a multi-beam RFI filter \multiTRAPUM{}\footnote{\href{https://github.com/mcbernadich/multiTRAPUM}{https://github.com/mcbernadich/multiTRAPUM} by Miquel Colom i Bernadich}. The latter is a wrapper of \texttt{rfifind} which removes signals detected in several beams with sufficient spatial separation. We switched to \pulsarX{}'s \texttt{filtool} \citep{Men2023} for the third pass as part of a TRAPUM pipeline upgrade, where we also applied a standard channel mask. 
A de-dispersion plan was computed with \presto{}'s \texttt{DDplan.py} script. The DM step sizes (i.e. the DM increment between each of the trial DMs) and sampling time are increased to match the growth of channel time smearing at higher DMs, to avoid unnecessarily high time resolution and thus spurious computation. We used the \dedisp{} library \citep{levin2012,Barsdell2012} for dedispersion over DMs up to 500\dmunits{} (where 1\,ms of combined intra-channel dispersion and de-dispersion step size smearing is reached) as part of the \peasoup{} suite, a GPU-based pulsar searching software \citep{PEASOUP,Morello2019,Sengar2022}.
The 2-hour long data were not corrected for acceleration before performing a Fast Fourier Transform (FFT) search using \peasoup{}. Thus, we searched for systems with no significant\footnote{Systems with low acceleration and long orbital periods can be detected, as long as the change in spin frequency is small compared to the frequency resolution of the FFT search \citep[see e.g.][]{Johnston1991}.} acceleration from a binary companion, with a minimum FFT spectrum Signal to Noise Ratio (S/N) of   8  and a maximum period of 10\,s. The 2-hour timeseries were zero-padded to the next nearest power of 2, $2^{26}$, resulting in a better resolved Fourier spectrum. Candidates were harmonically summed by \peasoup{} up to the eighth harmonic \citep{Taylor1969}.  The resulting spectral candidates were then sifted by \peasoup{}, i.e. clustered in DM, period and harmonics, in each beam independently. A list of common local RFI fluctuation frequencies was removed from the searches. This returned on average around 2000 candidates  in one coherent beam for one observation.

This was followed by multi-beam spatial sifting\footnote{\href{https://github.com/prajwalvp/candidate_filter}{https://github.com/prajwalvp/candidate\_filter} by Lars K\"{u}nkel}, which uses the expected spatial relationship of real sources to identify RFI \citep{Padmanabh2023}, after which less than 200 candidates remained in one beam for one observation.
We note that a recently discovered bug in the multi-beam filtering code over-removed candidates from an in-built list of RFI fluctuation frequencies. It is possible to retroactively check if a specific candidate has been removed in a list stored as a data product of this pipeline.
We only kept candidates with periods  longer than eight time samples (1.225\,ms), which typically removed a few candidates per beam in each observation. 
We folded the raw data with the parameters of the filtered FFT candidates using \pulsarX{}'s \psrfoldfil{} \citep{Men2023}, which downsamples the raw timeseries into folded data as follows: 64 time samples (bins) per candidate pulse profile (128 for signals with $P > 100$\,ms), each sub-integration of the observation in time is 20 seconds long, and the number of frequency channels is reduced to 64. \psrfoldfil{} cleans the full resolution data from RFI with \filtool{} before folding and applies the \clfd{} RFI removal software \citep{clfd} on the folded data. The folds were optimised to the highest S/N pulse profile over a period and period derivative range around the  \peasoup{}-detected topocentric spin frequency so that the maximum smearing is one pulse period over the whole observation time, and over a DM range so that the maximum dispersion delay is one pulse period over the whole bandwidth.

We then partially classified the folded candidates  with \pics{} \citep[a Pulsar Image-based Classification System based on Machine Learning]{PICS}. We used a minimum score of 10 per cent pulsar-like for both the original PALFA model \citep{PICS} and TRAPUM models \citep{Padmanabh2023}, as well as a minimum folded profile S/N of 7 to eliminate candidates.  This returned on average less than one  candidate per beam for each observation to classify visually. We used a Graphical User Interface candidate classifier tool, \candyjar{}\footnote{\href{https://github.com/vivekvenkris/CandyJar}{https://github.com/vivekvenkris/CandyJar} by Vivek Venkatraman Krishnan}, to classify the folded candidate plots by eye. Two or three viewers classified the candidates, most of which were consistent with RFI or noise, except for about five candidates per observation of each galaxy which were low confidence pulsar candidates. These were cross-checked between the first and second pass observations, and their discovery positions re-observed in the third pass. None were detected and classified as pulsar candidates more than once, and we hence discarded them as candidates arising from noise fluctuations.

\begin{figure*}
\centering
\includegraphics[width=0.6\linewidth]{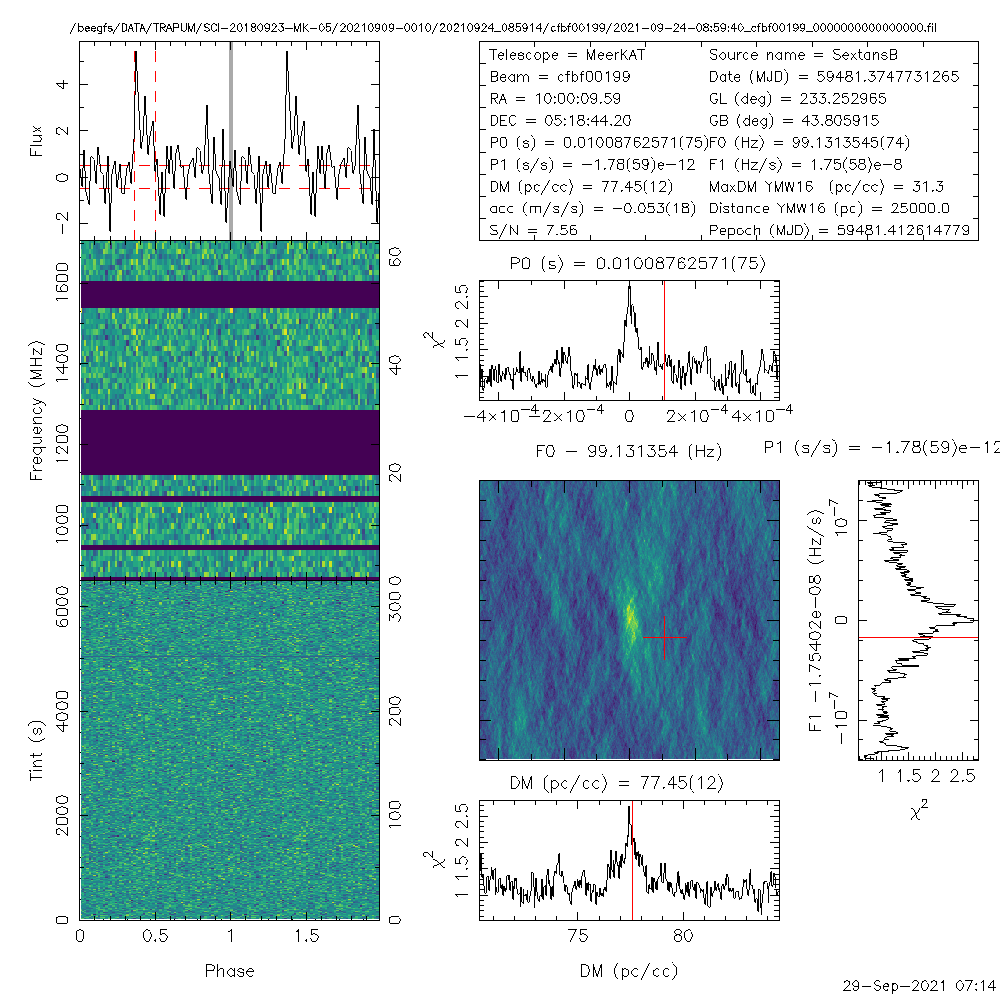}
\caption{An example of a low-confidence candidate from this survey, from the third pass observation of Sextans B. Details of the \pulsarX{} folding package can be found in \protect\cite{Men2023}.}
\label{fig:T2_example}
\end{figure*}

We also searched the full observations for single pulses with \transientX{} \citep{men2024} on the data downsampled by a factor of two in time to a sampling time of 306$\upmu$s, RFI cleaned by a \texttt{rfifind} channel mask and \pulsarX{}'s \texttt{filtool}. We used a 100\,ms maximum search width over DMs 0 to 5000\dmunits{} and a minimum single pulse S/N of 8. On average, several hundred candidates per observation were returned,  most of which came from the incoherent beam that is more affected by RFI due to its wide field of view. 
A visual inspection of  all the single pulse  candidates indicated they were mostly consistent with RFI or noise, except for less than five candidates per observation of each galaxy which were low-confidence candidates (see \autoref{fig:T2_example}), and one FRB discovery described in the next section. The low-confidence candidates were cross-checked between the three passes with no redetections, suggesting they could have arisen from noise fluctuations.  As the raw data were too voluminous to keep, we stored an RFI-cleaned (by \iqrm{}) downsampled (512 channels  and 1.225\,ms sampling time) version of the second and third passes. Apart from the offline processing using \transientX{}, a commensal search for transients was carried out on the incoherent beam and coherent beams using the MeerTRAP backend TUSE (Transient User Supplied Equipment). The capabilities and pipelines for MeerTRAP are described in \cite{Rajwade2022}.

\section{FRB\,20210924D}
\label{FRB}

\begin{figure}
\centering
\includegraphics[width=\columnwidth]{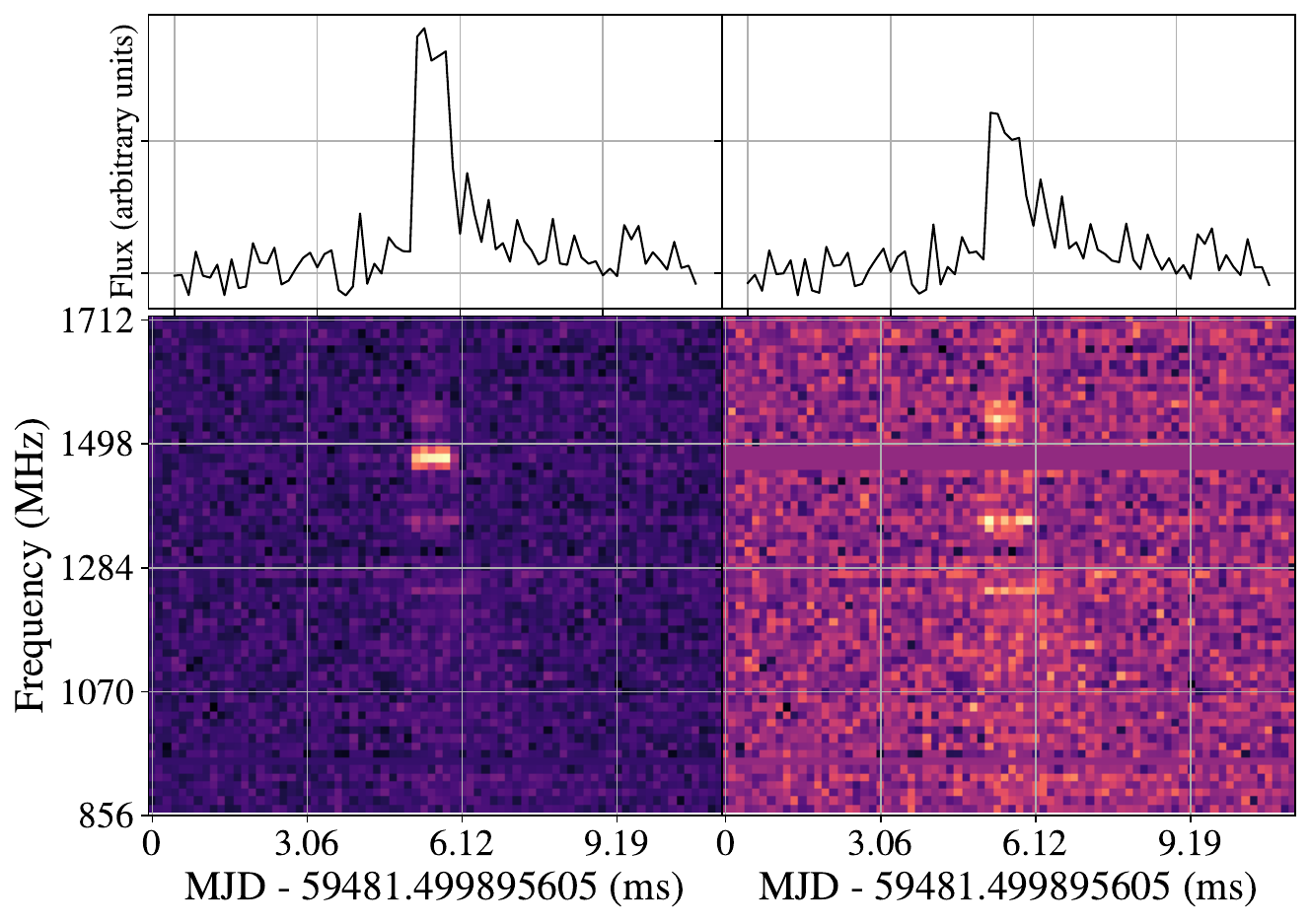} 
\caption[]{The detection of FRB\,20210924D in the incoherent beam of the third pass of Sextans A. In the bottom left plot, we show the dynamic spectrum of the FRB in the L-band of MeerKAT reduced to 64 subbands (from a recorded frequency resolution of 2048 channels) where RFI has been removed manually with \psrchive{}. The time resolution is the sampling time: 153\,$\upmu$s. In the bottom right plot, we show the same dynamic spectrum with the three subbands that contain the brightest narrowband emission of the FRB removed. This makes the fainter emission more perceptible. The top plots are the total flux added in frequency. \transientX{} evaluates the S/N of the burst in the left panel as 30.9 and its DM as $737.54\pm0.61$. }
\label{fig:FRB-IB}
\end{figure}

\begin{figure}
\centering
\includegraphics[width=\columnwidth]{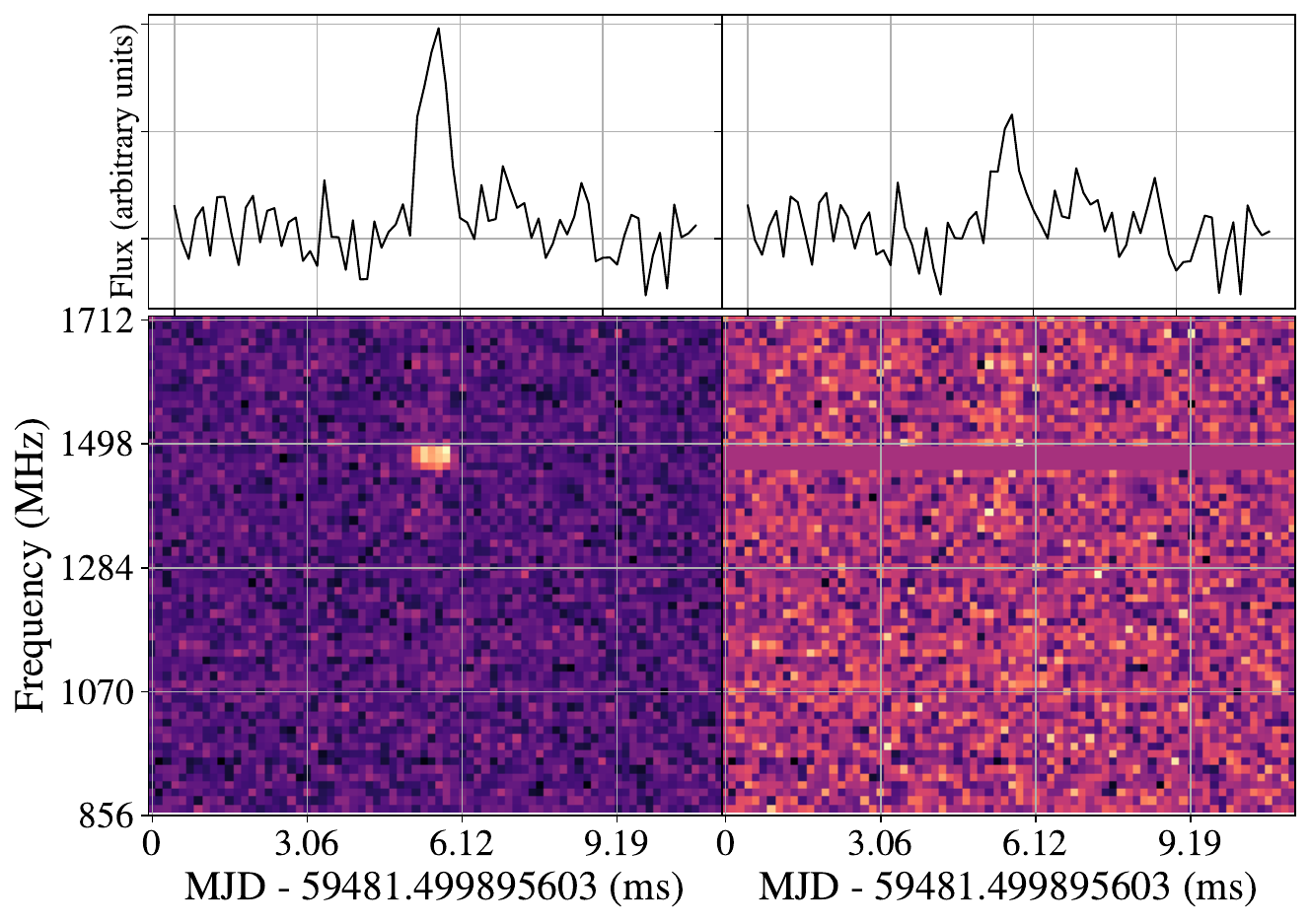} 
\caption[]{The strongest detection of FRB\,20210924D in a coherent beam of the third pass of Sextans A. The plot is as described in \autoref{fig:FRB-IB}. The S/N of the burst in the left panel (as evaluated by \transientX{}) is 13.1. }
\label{fig:FRB-CB}
\end{figure}

Two transient candidates from the third pass Sextans A observation matched in date (2021-09-24 11:59:50.98 UTC, MJD\,59481.49989563), width (about 1\,ms), and DM (737\dmunits{}). One was detected by the pipeline in a coherent beam (CB, number \texttt{000175}) at location RA(J2000) $=$ 10\h10\m59\fs71 Dec(J2000) $=$ $-$04\textdegree{}39\arcmin08\farcs20 with a S/N of 9.2 and the other in the incoherent beam (IB) centred on Sextans A with a S/N of 15.7. Inspection of the raw data revealed that bright narrowband emission from the transient had been partially masked by the \filtool{} RFI mitigation (as expected), lowering its S/N;  and that the pulse was present in many CBs with a lower S/N than in the IB\footnote{The S/N post-\filtool{} RFI cleaning was below the S/N$=$8 threshold for detection by our pipeline in all coherent beams but one. Therefore, the coherent beam with the strongest S/N was the only pipeline detection in the tiling.}.  This suggests that the transient occurred in the IB outside the coherent tiling, and was detected in the sidelobes of the CBs. The short-duration transient is consistent with an FRB due to its high DM. The narrowband nature of this FRB could be due to the beam response (e.g. if the FRB is located in an incoherent beam sidelobe, \citealt{Obrocka2015a}) or an intrinsic feature of the emission \citep{Xu2022}. Fainter emission is visible throughout the band (see \autoref{fig:FRB-IB}).  We found no published previous occurrences of this FRB and indeed no known FRBs within seven  degrees of Sextans A\footnote{Using the Transient Name Server, \href{https://www.wis-tns.org/}{https://www.wis-tns.org/}. The FRB report for this publication is available \href{https://www.wis-tns.org/object/20210924d}{here}.}. We show the incoherent beam detection in \autoref{fig:FRB-IB} and the strongest coherent beam detection in \autoref{fig:FRB-CB}. We used the software \textsc{scatfit} \citep{scatfit,Jankowski2023} to evaluate the properties of the FRB from the IB data, which we present in \autoref{tab:FRB-properties}. Scattering parameters could not be disentangled from dispersion effects, either due to weak scattering or the low S/N of the FRB at high frequencies.

\begin{table*}
\centering
\caption{The properties of FRB\,20210924D as fitted by \textsc{scatfit} on the incoherent beam data, where $W_{\text{eq}}$ is the equivalent width of the pulse and $W_{\text{50}}$ is the width of the burst at 50 per cent of the peak intensity.}
\label{tab:FRB-properties}

\begin{tabular}{ccccc}
\hline
Band name  & Central Frequency (MHz) & Fluence (Jy\,ms) & $W_{\text{eq}}$ (ms) & $W_{\text{50}}$ (ms) \\ \hline
Upper half & 1497.8                  & 0.83(4)          & 0.74(5)              & 0.65(4)       \\ 
Lower half & 1069.8                  & 1.3(2)           & 1.9(6)               & 1.6(3)        \\ 
Whole L-band & 1283.8                  & 0.90(7)          & 0.86(9)              & 0.65(8)       \\ \hline
\end{tabular}
\end{table*}

We attempted to localise the FRB by locating a transient source in the correlated visibility data obtained commensally in the third pass Sextans A observation (see \autoref{observations}, schedule block number \texttt{20210909-0010}). We performed this by subtracting all the persistent continuum sources from the visibility data and searching for residual emission in a single time cell corresponding to the time stamp when the FRB was detected in the IB and CBs. We limited our search to a small bandwidth of about 12\,MHz centred at 1468\,MHz since this frequency range contains the dominant emission in the incoherent and coherent beams. 

The narrow band visibility data, with a scan length of two hours and a time resolution of 8s, were calibrated following the standard procedure using the Common Astronomy Software Applications \citep[\textsc{casa};][]{CASA} package. Automated RFI flagging, usually part of the standard calibration workflow of interferometry data, was not performed on the Sextans A data to avoid inadvertent removal of the FRB signal. Manual RFI removal was performed instead. Next, we imaged the entire 2-hour observing block except for the single 8\,s time cell where the FRB signal might be present (we ensured the FRB was not dispersed beyond a single time cell). Ignoring the single time cell ensures that the FRB signal is not accidentally subtracted from the visibility data. We then generated a model of sources in the continuum image using the \textsc{pyBDSF} source finder \citep{pyBDSF}. We subtracted the generated model from the visibility data, which should only contain any remaining transient emission. We imaged the residual visibility data to produce a $5\degr \times 5\degr$ image. Searching the residual image visually and using the \textsc{pyBDSF} source finder yielded no detection, thus we were not able to identify the FRB in the image. The third pass image of Sextans A was generated from 62 antennas. Using the relationship between imaged and time-domain significance from \cite{Rajwade2022}, we expect  that the transient source would have a S/N of about 3--4 in an 8\,s-integrated image, which may explain this non-detection.

\begin{figure}
\centering
\includegraphics[width=\columnwidth]{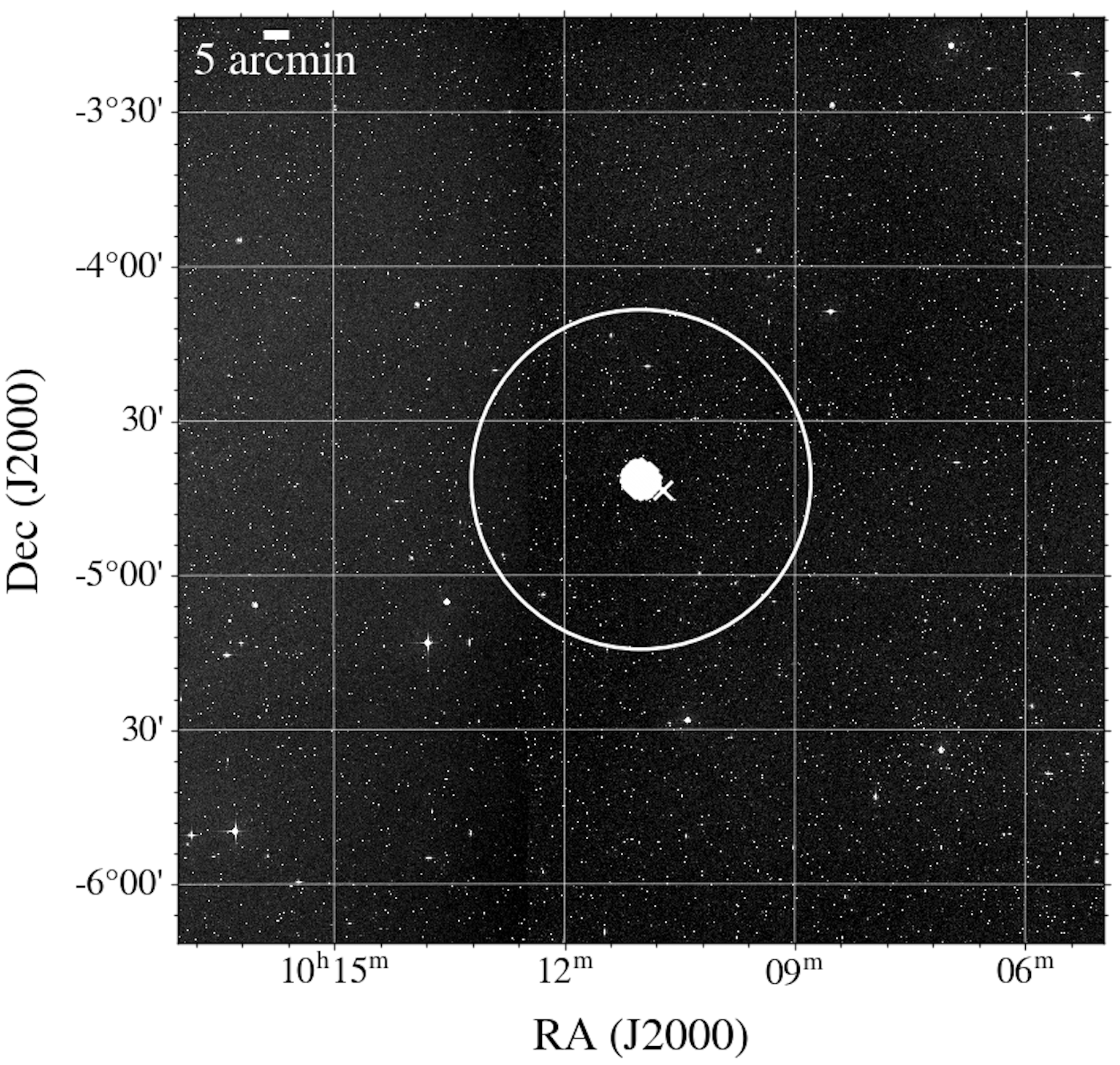}
\caption{The beams recorded in the third pass observation of Sextans A are displayed over a DSS image. The large circle is the MeerKAT incoherent beam size at half-maximum sensitivity at the centre frequency of L-band.  The Globular Cluster \citep{Beasley2019} is shown with a cross.  The observed MeerKAT full array coherent beams are shown at 50 per cent sensitivity as white ellipses, simulated with \mosaic{} \citep{Chen2021}. This figure was generated with the \href{http://aplpy.github.io/}{\aplpy{}} Python package. 
}
\label{fig:SextansA-thirdpass-IB}
\end{figure}

The sky position of the FRB is thus unconstrained. The MeerKAT incoherent beam has a FWHM of 1.1\,degrees at the centre frequency of L-band  and 0.9\,degrees  at the  highest frequency of the band \citep{Asad2021}, and the source could be far from the pointing centre, in a less sensitive area of the IB. We show the IB size relative to the coherent beam tiling in \autoref{fig:SextansA-thirdpass-IB}.

\section{Sensitivity limits}
\label{upperlimits}
\begin{table}
\centering
\caption{Survey sensitivity calculations parameters for our observations with the L-band receiver of MeerKAT \protect\citep{Bailes2020}.  The digitisation correction factor is taken from \protect\cite{digitisationfactors}.}
\label{tab:sensitivity_parameters_meerkat}
\resizebox{0.8\columnwidth}{!}{
\begin{tabular}{ll}
\textbf{Parameter}                            & \textbf{Value} \\ \hline
Bandwidth  $\upDelta\nu$ (MHz)                     & 856   \\ \hline
Centre frequency $\nu$ (MHz)               & 1284  \\ \hline
System temperature  $T_{\text{sys}}$  (K)             & 18    \\ \hline
Sky temperature $T_{\text{sky}}$ (K)                     & 6.5   \\ \hline
Gain $G$ (60 coherent antennas, K\,Jy$^{-1}$)         & 2.625 \\ \hline
Gain  $G$ (62 incoherent antennas, K\,Jy$^{-1}$)      & 0.344  \\ \hline
Number of frequency channels         & 2048  \\ \hline
Sampling time ($\upmu$s)               & 153   \\ \hline
Number of polarisations recorded $n_{\text{pol}}$    & 2     \\ \hline
Digitisation correction factor $\beta$ (8-bit)      & 1.0     \\ \hline
\end{tabular}}
\end{table}

\begin{table}
\centering
\caption{Survey sensitivity calculations parameters. The sky temperature was retrieved from \protect\cite{skytemperature} using \pygsm{} \protect\citep{pygsm}. }
\label{tab:other_sensitivity_parameters}
\resizebox{0.8\columnwidth}{!}{
\begin{tabular}{ll}
\textbf{Parameter}                            & \textbf{Value} \\ \hline
Assumed pulsar duty cycle $\delta$ (per cent) & 2.5     \\ \hline
Minimum spectral S/N searched to    S/N$_{\text{min}}$       & 8     \\ \hline
FFT efficiency factor $\epsilon$ & 0.7 \\ \hline
Observation time  $t_{\text{obs}}$ & 7200\,s \\ \hline
\end{tabular}}
\end{table}

We calculate radio flux density sensitivity limits  using the the radiometer equation (\autoref{eq:radiometer}) applied to pulsar observations \citep[p. 265]{handbook} with the usual simplification of a constant sensitivity across the bandwidth and a flat pulsar spectrum:

\begin{equation}
S_{\text{min}}  = \frac{\text{S/N}_{\text{min}}  \times  ( T_{\text{sys}}  + T_{\text{sky}})  \times  \beta }{\epsilon \times G \times  \sqrt{n_{\text{pol}}  \times  t_{\text{obs}}  \times  \upDelta\nu }}  \times  \sqrt{\frac{\delta}{1 - \delta}} \ \, .
\label{eq:radiometer}
\end{equation} 

\begin{figure}
\centering
\includegraphics[width=1.1\columnwidth]{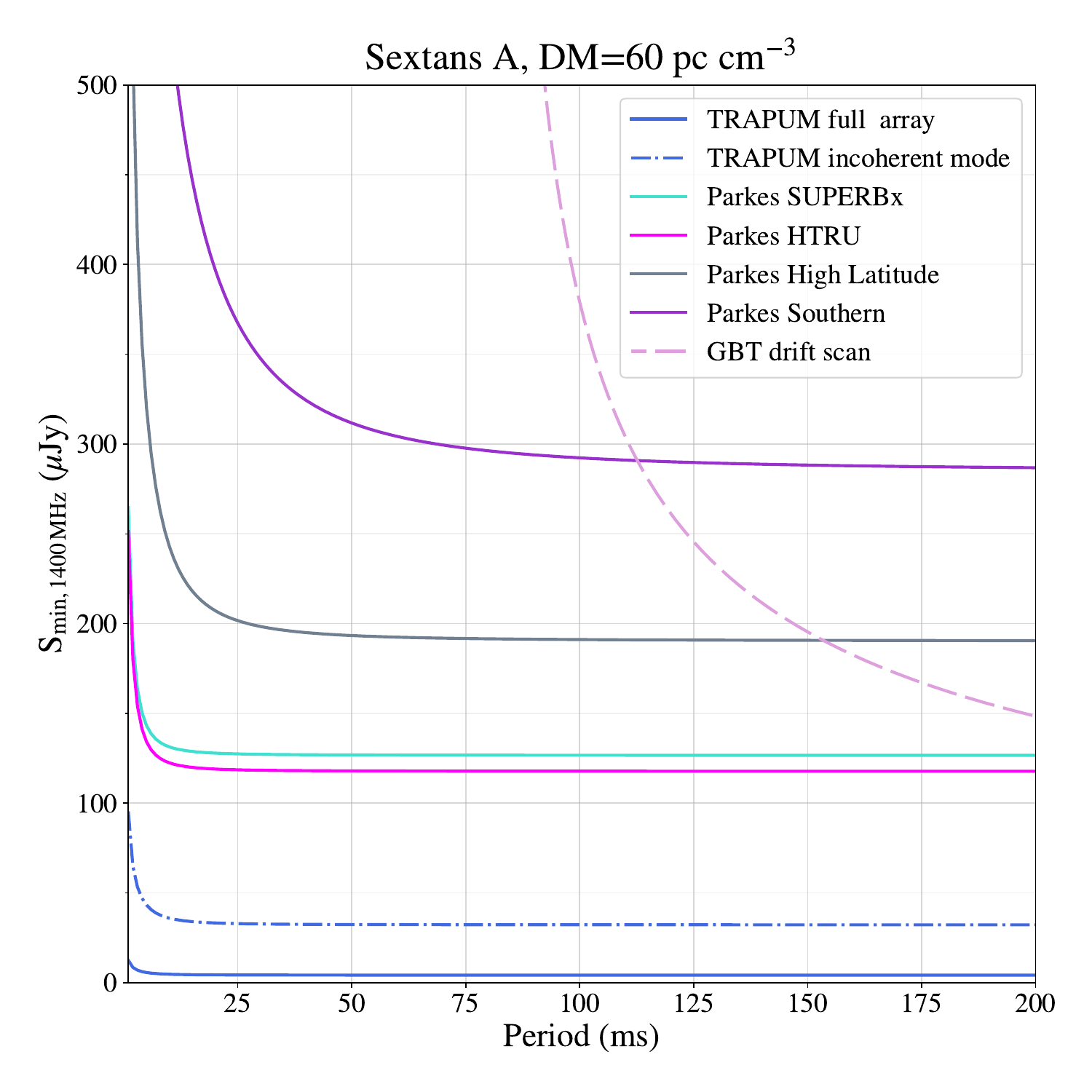} 
\caption[caption:SextansA-survey-sensitivities]{Flux density sensitivity limits as a function of pulsar period for all pulsar surveys that have observed Sextans A. The maximum period searched by our survey is 10\,s: we have shown here the shortest periods in order to indicate where our sensitivity deviates from being quasi-constant as a function of period.}
\label{fig:SextansA-survey-sensitivities}
\end{figure}

We use the parameters detailed in  \autoref{tab:sensitivity_parameters_meerkat} for MeerKAT in a full array configuration with 60 coherent antennas at L-band at the centre of the pointing (primary beam location), and at the centre of a coherent beam (where it has maximum sensitivity). Other parameters used in our sensitivity calculations are given in \autoref{tab:other_sensitivity_parameters}. We use a FFT efficiency factor $\epsilon = 0.7$ to perform a spectral to folded S/N conversion \citep{Morello2019}. Thus, a minimum S/N cut of  8   in the FFT corresponds to a folded S/N of approximately 12. For a pulsar with a rotation period of 100\,ms, this yields a flux density limit of $S_{\text{1284\,MHz}} =$ 4.8\,$\upmu$Jy. 
At shorter periods, pulse widening due to  sampling time, dispersion smearing in a frequency channel \citep[p. 109]{handbook}, and dispersion scattering \citep{Lewandowski2015} are introduced. Due to these effects, at shorter periods, the duty cycle $\delta$ departs from our assumed intrinsic 2.5\,per cent value\footnote{This value of $\delta$ is the median intrinsic duty cycle from the ATNF pulsar catalogue, for radio-emitting pulsars with a measured pulse width,  excluding GC pulsars, MSPs, Rapidly Rotating Radio Transients, magnetars and binary pulsars.} (\autoref{fig:SextansA-survey-sensitivities}), increasing the flux density limit to 14\,$\upmu$Jy at 1.225\,ms.  The sensitivity reduction for a coherent beam placed at the edge of the optical extent of the dwarf galaxy is less than one per cent (using a primary beam sensitivity model based on \citealt{Asad2021}). We can also provide a limit for the Sextans A globular cluster in the first pass observation, accounting for factors that caused the sensitivity to be decreased, i.e. the reduced number of antennas,  the position in the incoherent beam and in the coherent beam: 7, 1 and 50 per cent sensitivity reductions respectively. For a pulsar with a rotation period of 1.225\,ms (including pulse broadening) this yields a flux density limit of $S_{\text{GC,1284\,MHz,MSP}} =$ 23\,$\upmu$Jy for the Sextans A globular cluster. 

In \autoref{fig:SextansA-survey-sensitivities}, we show the sensitivities of all pulsar surveys that have observed Sextans A: this work, the Parkes Southern pulsar survey \citep{Manchester1996}, the Parkes high-latitude pulsar survey \citep{Burgay2006}, the High Time Resolution Universe Pulsar Survey \citep[HTRU]{Keith2010},  the Green Bank telescope 350\,Mhz drift-scan survey \citep{Boyles2013,Lynch2013}, and the Parkes Survey for Pulsars and Extragalactic Radio Bursts \citep[SUPERB]{Keane2018}. 
Again, we use the the radiometer equation applied to pulsar observations (\autoref{eq:radiometer}) to calculate their flux density sensitivity limits. 
We find the values of the parameters listed in  \autoref{tab:sensitivity_parameters_meerkat} for the surveys in their respective papers, including their minimum S/N, except for duty cycle which we set at 2.5 per cent, and sky temperature. We compute the latter at each survey's central frequency with \pygsm{} \citep{pygsm}. We assume the gain of Murriyang to be 0.735 K\,Jy$^{-1}$ in all cases, using the boresight value \citep{Manchester2001}.   The digitisation correction factors are taken from \cite{digitisationfactors}: 1.0 for 8-bit digitisation, 1.25 for 1-bit, and 1.06 for 2-bit.
We rescale all surveys to 1400 MHz assuming a -1.60 pulsar spectral index \citep{Jankowski2018}. We note that the Parkes Southern pulsar survey and the Green Bank telescope drift-scan survey have central frequencies of 436 and 350\,MHz respectively. This means that the spectral index conversion is speculative as there is a range of pulsar spectral  indices that may vary the resulting sensitivity. Other surveys compared were all conducted at L-band, close to a central frequency of 1400\,MHz. We use the highest DM that has been searched by all the surveys, 60\dmunits{}.
We again take into account pulse widening effects as detailed previously. Temperature contributions from the ground and the atmosphere, scintillation effects, effective bandwidth (due to RFI masking and band characteristics), de-dispersion step size, and harmonic summing contributions  are not taken into account. 
As shown in \autoref{fig:SextansA-survey-sensitivities}, our radio flux density limit rescaled at 1400\,MHz for the mean spectral index is $S_{\text{1400\,MHz}} =$ 4.2$\pm$0.2\,$\upmu$Jy at the centre of the pointing, for a pulsar with a rotation period of 100\,ms. This 1400\,MHz flux density limit is  28 times more sensitive than the best previous flux density limit from \cite{Keith2010}. Sextans B has been observed by the same surveys except HTRU, and has a similar sky temperature, and thus similar flux limits.

\begin{figure}
\centering
\includegraphics[width=1.1\columnwidth]{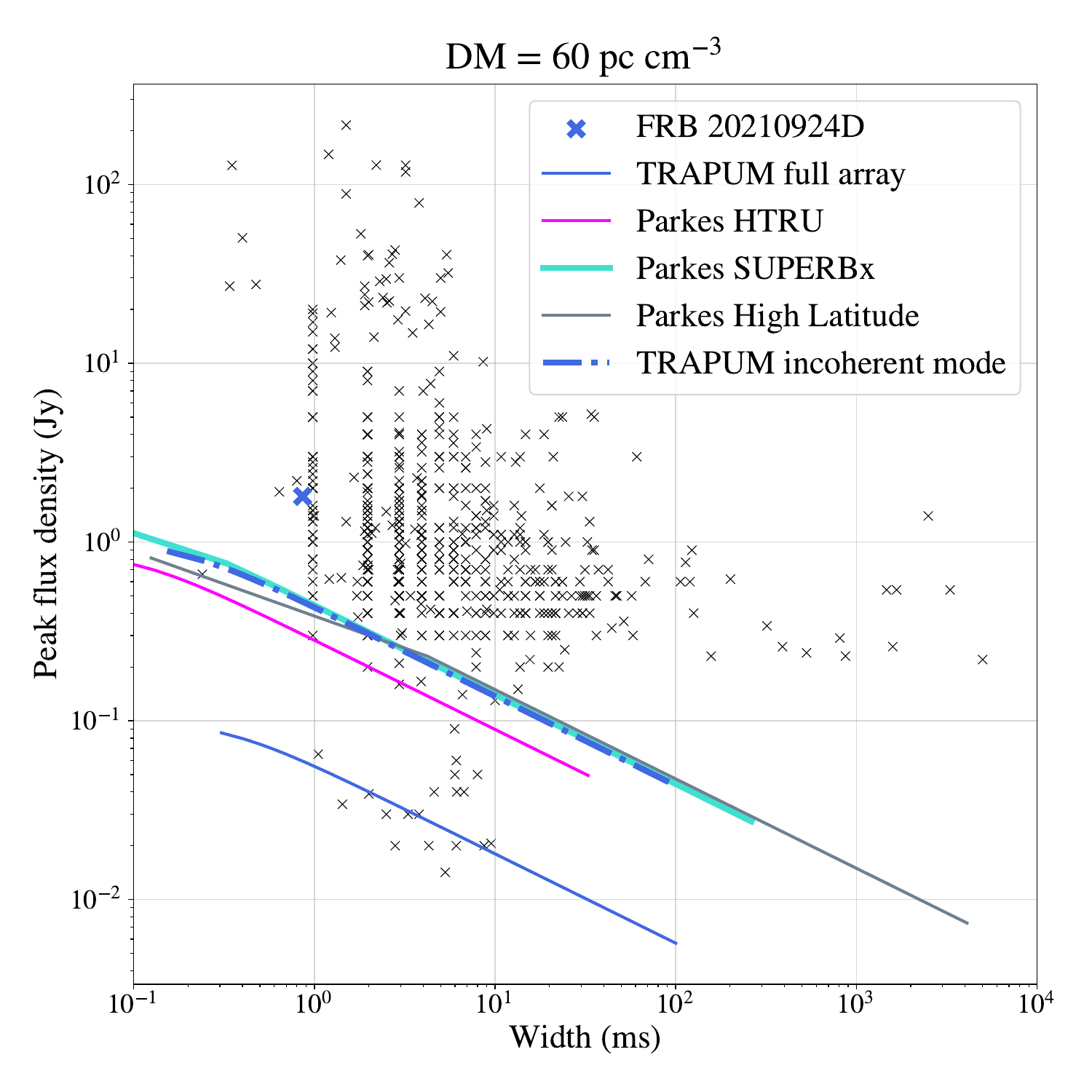} 
\caption[caption:transient-sensitivity]{Flux density sensitivity limits of all L-band surveys that have performed a targeted search for transient dispersed pulses from Sextans A. Known FRBs are plotted as black crosses, from the \href{https://www.herta-experiment.org/frbstats/catalogue}{FRBSTATS} catalogue. We note that the catalogued FRB flux densities are calculated with varying methods  and should be treated as an approximation. Indeed, the true flux density of FRB\,20210924D is unknown due to its unknown position in the incoherent beam sensitivity response. The quantisation of FRB values is due to rounding and sampling time limitations.}
\label{fig:transient-sensitivity}
\end{figure}

A limiting single pulse flux density sensitivity plot of transient searches of Sextans A is shown in \autoref{fig:transient-sensitivity}. All L-band surveys mentioned earlier except Parkes Southern have searched for dispersed single pulses: the Parkes high-latitude pulsar survey \citep{Rane2016}, the High Time Resolution Universe Pulsar Survey \citep{Burke-Spolaor2011},   and the Parkes Survey for Pulsars and Extragalactic Radio Bursts \citep[SUPERB]{Keane2018}. We do not compare with the Green Bank telescope 350\,MHz drift-scan survey \citep{Boyles2013,Lynch2013} or CHIME \citep{CHIME} due to the difference in observed frequency. We find the values of the parameters listed in  \autoref{tab:sensitivity_parameters_meerkat} for the surveys in their respective papers, and the same assumptions as before. We also use the minimum single pulse S/N and the single pulse widths  covered by their search (starting from the sampling time). We compute the sky temperature at each survey's central frequency with \pygsm{}. We use these values in the following  equation from \cite{Cordes2003} to calculate pulse flux density:

\begin{gather*}
S_{\text{min,pulse}}  = \frac{ S/N_{\text{min}}  \times  (T_{\text{sys}}  + T_{\text{sky}})  \times  \upbeta }{G\times \sqrt{n_{\text{pol}} \times \upDelta \nu \times w_{\text{broadened}} } } \ \, ,
\end{gather*}

\noindent where $w_{\text{broadened}}$ is the observed width of the transient signal. We calculate the broadened pulse width by taking into account sampling time and dispersion smearing. We assume a DM of 60\dmunits{}. The surveys have all searched for single pulses to at least a 1000\dmunits{}, thus a repeat FRB from the source of FRB\,20210924D should have been detected if it had repeated during their observations (DM$=737$\dmunits{}).  The limiting peak flux density for the single pulse search down to S/N$=$8 is $S_{\text{pulse,1284\,MHz}} =$ 56\,mJy\,ms for an intrinsic pulse width of 1\,ms at 60\dmunits{}. The limiting fluence for the Sextans A globular cluster is thus $S_{\text{GC,pulse,1284\,MHz}} =$ 89\,mJy\,ms, for an intrinsic pulse width of 1\,ms at 60\dmunits{}.  Again, this greatly improves the previous upper limit  of 284\,mJy\,ms  for an intrinsic pulse width of 1\,ms at 1352 MHz and 60\dmunits{} \citep{Burke-Spolaor2011}.

\section{Discussion and conclusions}
\label{discussion}

For a pulsar with a rotation period of 100\,ms, our flux density sensitivity at 1400\,MHz translates to a radio pseudo-luminosity\footnote{This is an approximation of the pulsar luminosity using a flux density and distance approximation.} limit of $L_{\text{pseudo,1400\,MHz}}= S_{\text{1400\,MHz}} \times D^{2} = $ 7.9$\pm$0.4\,Jy\,kpc$^{2}$,  assuming an approximate distance to the galaxies of $D=$ 1375\,kpc \citep{Tully2013,Bellazzini2014}, and a power law radio spectral index of -1.60$\pm$0.54. According to the ATNF pulsar database \citep{ATNF}, the highest pseudo-luminosity for a galactic pulsar is held by B1641$-$45 (also known as J1644$-$4559) with 6.1\,Jy\,kpc$^{2}$ \citep{Komesaroff1973,Jankowski2018}. The highest pseudo-luminosity for an extragalactic pulsar is J0523$-$7125 in the LMC with 2.5\,Jy\,kpc$^{2}$ \citep{Wang2022}. If the extreme  high end of the pulsar luminosity distribution of the galaxies Sextans A and B extends slightly further than the Milky Way's and above our limit, no such bright pulsars are beamed in our direction and were unobscured during our observations (e.g. due to excessive DM, scattering, or eclipsing), as we have observed the entire dwarf galaxies. However, the beaming fraction of pulsars is estimated to be around 20\,per cent \citep{Taylor1977}, though newer surveys may alter this value \citep[e.g.][]{Turner2024}.

\begin{figure}
\centering
\includegraphics[width=\columnwidth]{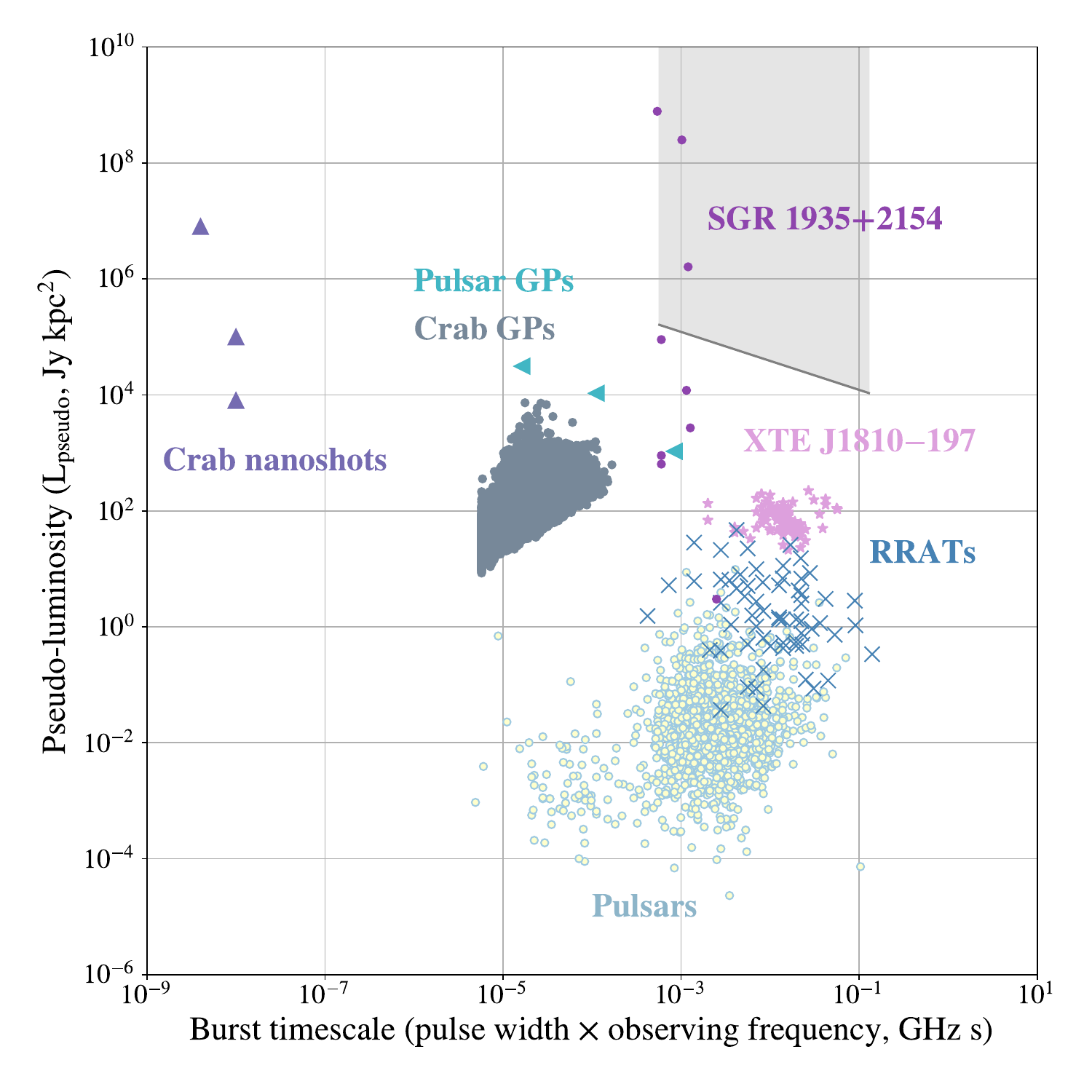} 
\caption[caption:phasespace]{Pseudo-luminosity versus burst timescale of radio bursts from neutron stars. The sensitivity of our survey of the Sextans A galaxy is shaded in grey. The Sextans B observations yield similar results. Giant pulses (GPs) and regular pulses from pulsars and Rapidly Rotating Radio Transients (RRATS), as well as bursts from two magnetars, XTE\,J1810$-$197 and SGR\,1935$+$2154, are depicted. Data and figure courtesy of Manisha Caleb, adapted from \cite{Driessen2023}, originally from \cite{pietka2015}. The source code of this figure is available  on 
 this \href{https://github.com/FRBs/Transient_Phase_Space}{GitHub repository}.}
\label{fig:phasespace}
\end{figure}

As shown in \autoref{fig:phasespace}, we could not detect any single pulses from known Rapidly Rotating Radio Transients or giant pulses from known pulsars  if they were situated at the distance of the Sextans galaxies and beamed in our direction -- including the extragalactic Crab analogue  PSR\,B0540$-$6919 \citep{Seward1984,Johnston2003}. We could not detect very short timescale pulses either as our time resolution was too large in the transient search (306\,$\upmu$s). However, we could detect FRB-like single pulses, similar to those seen from the Galactic magnetar SGR 1935+2154 \citep{Bochenek2020b,CHIME2020b}. As stated in \autoref{introduction}, the presence of magnetars in Sextans A and B is made more likely by the recent star formation in these galaxies.  From our search, we conclude that no FRBs or FRB-like events from Sextans A or B were beamed in our direction at the time of observing. We did detect a new FRB in the incoherent and a number of (widely spaced) coherent beams of one of the Sextans A observations, but we were not able to constrain its sky position. We believe it is not associated with the dwarf galaxy due to its S/N being strongest in the incoherent beam. All detections in the coherent beams, which were covering the entire galaxy, are consistent with being detected via sidelobes. We can assume the sum of the Milky Way (see \autoref{introduction}) and Intergalactic Medium DM contributions is of the order of 50\dmunits{} at the redshift of Sextans A ($z_{\text{Sextans}} \simeq 3 \times 10^{-4}$), using the Macquart relation \citep{Macquart2020}. We are thus left with 687\dmunits{} to account for, which are unlikely to be attributed to the tenuous Sextans A galaxy content, but rather a much more distant galaxy host at an expected redshift of $z_{\text{host}}\simeq0.6$. This new transient, FRB\,20210924D, is not, to our knowledge, a known repeater despite being visible to CHIME \citep{CHIME}.

\section*{Acknowledgements}
The MeerKAT telescope is operated by the South African Radio Astronomy Observatory, which is a facility of the National Research Foundation, an agency of the Department of Science and Innovation. SARAO acknowledges the ongoing advice and calibration of GPS systems by the National Metrology Institute of South Africa (NMISA) and the time space reference systems department of the Paris Observatory.

TRAPUM observations used the FBFUSE and APSUSE computing clusters for data acquisition, storage and analysis. These clusters were funded and installed by the Max-Planck-Institut für Radioastronomie and the Max-PlanckGesellschaft.

EC acknowledges funding from the United Kingdom's Research and Innovation Science and Technology Facilities Council (STFC) Doctoral Training Partnership, project reference 2487536. 

AP and MB acknowledge that part of this work has been funded using resources from the INAF Large Grant 2022 `GCjewels' (P.I. Andrea Possenti) approved with the Presidential Decree 30/2022.

EB, MK, PVP and VVK acknowledge continuing support from the Max Planck society. 

RPB acknowledges support from the ERC under the European Union's Horizon 2020 research and innovation programme (grant agreement No. 715051; Spiders).

We acknowledge the use of NASA's \textit{SkyView} facility (\href{http://skyview.gsfc.nasa.gov}{http://skyview.gsfc.nasa.gov}) located at NASA Goddard Space Flight Center to access the Digitized Sky Survey images of Sextans A and B. The Digitized Sky Survey was produced at the Space Telescope Science Institute under U.S. Government grant NAG W-2166. The images of these surveys are based on photographic data obtained using the Oschin Schmidt Telescope on Palomar Mountain and the UK Schmidt Telescope. The plates were processed into the present compressed digital form with the permission of these institutions.

 This research made use of APLpy, an open-source plotting package for Python \citep{Robitaille2012}. 

This research has made use of the SIMBAD database, operated at CDS, Strasbourg, France \citep{SIMBAD}. This research has made use of NASA’s \href{https://ui.adsabs.harvard.edu/}{Astrophysics Data System} Bibliographic Services and the \href{https://www.herta-experiment.org/frbstats/catalogue}{FRBSTATS} catalogue. 

This paper has made use of the ATNF pulsar catalogue version 1.69.

\section*{Data Availability}
The data underlying this article will be shared upon reasonable request to the TRAPUM collaboration. The first and second pass observations were stored in a reduced resolution format only, cleaned by \iqrm{}. The resolution was decreased to 512 channels 
and  a 1.225\,ms sampling time. The raw data of all pointings were deleted immediately after processing as they were too large to store. All secondary processing data products were retained for possible future inspection, including candidates which can be readily shared with interested observers.

\bibliographystyle{mnras}
\bibliography{main.bbl}

\bsp	
\label{lastpage}
\end{document}